# Quantum Ground States as Equilibrium Particle-Vacuum Interaction States

Harold E. Puthoff

**Abstract** A remarkable feature of atomic ground states is that they are observed to be radiationless in nature, despite (from a classical viewpoint) typically involving charged particles in accelerated motions. The simple hydrogen atom is a case in point. This universal ground-state characteristic is shown to derive from particle-vacuum interactions in which a dynamic equilibrium is established between radiation emission due to particle acceleration, and compensatory absorption from the zero-point fluctuations of the vacuum electromagnetic field [1]. The result is a net radiationless ground state. This principle constitutes an overarching constraint that delineates an important feature of quantum ground states.

**Keywords** Quantum ground states · Vacuum fluctuations · Particle-vacuum interaction states · Zero-point fluctuations · Harmonic oscillator quantum ground state

## 1 Introduction

One of the apparent paradoxes of quantum theory that students often query is the radiationless nature of atomic ground states. The paradox lies in the fact that radiation that might be anticipated from accelerated charged-particle motions in atomic ground states is not observed to occur. In the hydrogen atom, for example, the orbiting electron does not radiate its energy away and spiral into the nucleus. The fact that during decades of successful application of quantum theory we have come to take for granted the radiationless feature of these special "bottom-rung" stationary states does not in any way detract from this remarkable property. Fortunately, a rapprochement between classical and quantum viewpoints is possible.

___________________________

H. Puthoff

Institute for Advanced Studies at Austin, 11855 Research Blvd., Austin, TX 78759, USA

e-mail: puthoff@earthtech.org

When addressed in analytical detail it becomes clear wherein the resolution to this apparent paradox lies. It is that one must properly take into account how charged-particle ground-state motions interact with the vacuum, specifically the zero-point fluctuations of the vacuum electromagnetic field. Although such considerations are not usually invoked in the day-to-day application of quantum theory to ground-state specification, the argument that follows clarifies that the standard formalism leading to radiationless ground states has its genesis in the dynamics of underlying particle-vacuum interactions, and that the vacuum field is in fact formally necessary for the stability of atoms in quantum theory. As summarized in an earlier paper addressing spontaneous emission processes: "The crucial role of the vacuum fluctuations emerges in the ground state of matter. The stability of the ground state (i.e., the fact that it does not radiate) is purely a quantum effect which is due to the vacuum fluctuations [2]."

It is sufficient for our purposes to treat such problems semiclassically on the basis of point particles interacting with a random, classical radiation field whose spectral characteristics are those of the known quantum vacuum zero-point fluctuation (ZPF) distribution. This approach, known as Stochastic Electrodynamics (SED), takes advantage of the fact that SED derivations of vacuum-fluctuation-driven phenomena based on multipole/radiation-field interactions parallel closely Heisenberg-picture derivations in standard QED [3,4]. Specifically, in such cases calculations in SED are analogous to QED calculations with a symmetric ordering of photon creation and annihilation operators [5]. Before considering application on the basis of a general formalism, let us apply the central argument in detail to the simple one-dimensional harmonic oscillator.

**2 Nonrelativistic Harmonic Oscillator**

For a one-dimensional harmonic oscillator of natural frequency $\omega_0$ located at the origin $\mathbf{r} = 0$ and immersed in the vacuum ZPF field, the (nonrelativistic) Abraham-Lorentz equation of motion for a particle of mass $m$ and charge $q$, including radiation damping, is given by [6]

$$m\ddot{x} + m\omega_0^2 x = \left(\frac{q^2}{6\pi\varepsilon_0 c^3}\right)\dddot{x} + qE_x^{zp}, \qquad (1)$$

where $E_x^{zp}$ is the $x$ component of the vacuum ZPF electric field and here we neglect the force contribution from the magnetic field (see Section 3, however).

The required expression for the electric field is obtained from the electromagnetic vacuum ZPF distribution whose spectral energy density is given by the Lorentz-invariant expression [7,8]

$$\rho(\omega)d\omega = \frac{\hbar\omega^3}{2\pi^2 c^3}d\omega, \qquad (2)$$

which corresponds to an energy $\hbar\omega/2$ per normal mode. In the SED ansatz the Fourier composition underlying this spectrum can be written as a sum of plane waves

$$\mathbf{E}^{zp} = \text{Re}\sum_{\sigma=1}^{2}\int d^3k\,\hat{\varepsilon}\sqrt{\frac{\hbar\omega}{8\pi^3\varepsilon_0}}e^{i\mathbf{k}\cdot\mathbf{r}-i\omega t+i\theta(\mathbf{k},\sigma)}.\qquad(3)$$

A similar expression for the magnetic field is obtained by replacing $\mathbf{E}^{zp}$ by $\mathbf{H}^{zp}$, $\hat{\varepsilon}$ by $(\hat{\mathbf{k}}\times\hat{\varepsilon},)$ and $\varepsilon_0$ by $\mu_0$. In these expressions Re denotes 'Real part of,' $\sigma=1,2$ denote orthogonal polarizations, $\hat{\varepsilon}$ and $\hat{\mathbf{k}}$ are orthogonal unit vectors in the direction of the electric field polarization and wave propagation vectors, respectively, $\theta(\mathbf{k},\sigma)$ are random phases distributed uniformly in the interval $0$ to $2\pi$ (independently distributed for each $\mathbf{k},\sigma$), and $\omega=kc$.

Substitution of Eq. (3) into Eq. (1) leads to the following expressions for position, velocity and acceleration:

$$x = \frac{q}{m}\text{Re}\sum_{\sigma=1}^{2}\int d^3k\,(\hat{\varepsilon}\cdot\hat{\mathbf{x}})\sqrt{\frac{\hbar\omega}{8\pi^3\varepsilon_0}}\frac{1}{D}e^{i\mathbf{k}\cdot\mathbf{r}-i\omega t+i\theta(\mathbf{k},\sigma)},\qquad(4)$$

$$v = \dot{x} = \frac{q}{m}\text{Re}\sum_{\sigma=1}^{2}\int d^3k\,(\hat{\varepsilon}\cdot\hat{\mathbf{x}})\sqrt{\frac{\hbar\omega}{8\pi^3\varepsilon_0}}\left(-\frac{i\omega}{D}\right)e^{i\mathbf{k}\cdot\mathbf{r}-i\omega t+i\theta(\mathbf{k},\sigma)},\qquad(5)$$

$$a = \ddot{x} = \frac{q}{m}\text{Re}\sum_{\sigma=1}^{2}\int d^3k\,(\hat{\varepsilon}\cdot\hat{\mathbf{x}})\sqrt{\frac{\hbar\omega}{8\pi^3\varepsilon_0}}\left(-\frac{\omega^2}{D}\right)e^{i\mathbf{k}\cdot\mathbf{r}-i\omega t+i\theta(\mathbf{k},\sigma)},\qquad(6)$$

where

$$D = -\omega^2 + \omega_0^2 - i\Gamma\omega^3,\qquad(7)$$

$$\Gamma = \frac{q^2}{6\pi\varepsilon_0 mc^3}.\qquad(8)$$

Now, for bounded, steady-state motion, we assume stationary expectation values for the mean-square position variable $x$ and its time derivatives. Thus,

$$\langle x^2\rangle = \langle x\cdot x\rangle = \frac{q^2}{m^2}\sum_{\sigma=1}^{2}\sum_{\sigma'=1}^{2}\int d^3k\int d^3k'\,(\hat{\varepsilon}\cdot\hat{\mathbf{x}})(\hat{\varepsilon}'\cdot\mathbf{x})\sqrt{\frac{\hbar\omega}{8\pi^3\varepsilon_0}}\sqrt{\frac{\hbar\omega'}{8\pi^3\varepsilon_0}}\frac{1}{DD'^*}$$

$$\times \frac{1}{2} \text{Re} \langle \exp[i(\mathbf{k} - \mathbf{k}') \cdot \mathbf{r} - i(\omega - \omega')t + i\theta(\mathbf{k}, \sigma) - i\theta(\mathbf{k}', \sigma')] \rangle, \quad (9)$$

where use of the complex conjugate and the notation $(1/2)\text{Re}$ derive from use of exponential notation. With $\int d^3k \to \int d\Omega_k \int dk k^2$, and averaging over random phases

$$\text{Re} \langle \exp[i(\mathbf{k} - \mathbf{k}') \cdot \mathbf{r} - i(\omega - \omega')t + i\theta(\mathbf{k}, \sigma) - i\theta(\mathbf{k}', \sigma')] \rangle = \delta_{\sigma\sigma'} \delta_{\omega\omega'} \delta^3(\mathbf{k} - \mathbf{k}'). \quad (10)$$

Equation (9) can therefore be simplified to

$$\langle x^2 \rangle = \frac{q^2}{2m^2} \int d\Omega_k \left[ \sum_{\sigma=1}^{2} (\hat{\varepsilon} \cdot \hat{\mathbf{x}})^2 \right] \int dk k^2 \left[ \frac{\hbar \omega}{8\pi^3 \varepsilon_0} \right] \frac{1}{DD^*}. \quad (11)$$

We further note that, with the sum over polarizations given by

$$\sum_{\sigma=1}^{2} [\hat{\varepsilon}(\mathbf{k}, \sigma) \cdot \hat{\mathbf{x}}_i][\hat{\varepsilon}(\mathbf{k}, \sigma) \cdot \hat{\mathbf{x}}_j] = \delta_{ij} - (\hat{\mathbf{k}} \cdot \hat{\mathbf{x}}_i)(\hat{\mathbf{k}} \cdot \hat{\mathbf{x}}_j), \quad (12)$$

the angular integration in $k$ takes the form

$$\int d\Omega_k \left[ \sum_{\sigma=1}^{2} (\hat{\varepsilon} \cdot \hat{\mathbf{x}})^2 \right] = \int d\Omega_k \left[ 1 - (\hat{\mathbf{k}} \cdot \hat{\mathbf{x}})^2 \right] = \frac{8\pi}{3}. \quad (13)$$

Substitution of Eq. (13) into Eq. (11), and a change of variables to $\omega = kc$, then leads to

$$\langle x^2 \rangle = \frac{q^2 \hbar}{6\pi^2 \varepsilon_0 m^2 c^3} \int_0^\infty \frac{\omega^3 d\omega}{DD^*} = \frac{q^2 \hbar}{6\pi^2 \varepsilon_0 m^2 c^3} \int_0^\infty \frac{\omega^3 d\omega}{\left(-\omega^2 + \omega_0^2\right)^2 + \Gamma^2 \omega^6}. \quad (14)$$

Due to the smallness of $\Gamma$ for charge-to-mass ratios of interest the integrand in Eq. (14) is sharply peaked around $\omega = \omega_0$. We therefore invoke the standard resonance approximation, extending the limits of integration and replacing $\omega$ by $\omega_0$ in all but the difference term. This yields, with substitution of the definition of $\Gamma$ from Eq. (8), the mean square fluctuation in position as

$$\langle x^2 \rangle = \frac{\hbar}{2m\omega_0} \int_{-\infty}^{\infty} \frac{1}{\pi} \frac{\left(\Gamma \omega_0^2 / 2\right) d\omega}{(\omega_0 - \omega)^2 + \left(\Gamma \omega_0^2 / 2\right)^2} = \frac{\hbar}{2m\omega_0}, \quad (15)$$

since the (Lorentzian lineshape) integral is unity.

Calculation of the mean square fluctuation in velocity $\langle v^2 \rangle = \langle \dot{x}^2 \rangle$ follows as above, except that $(1/DD^*) \rightarrow (-i\omega/D)(i\omega/D^*) = (\omega^2/DD^*)$, yielding

$$\langle v^2 \rangle = \frac{\hbar \omega_0}{2m}. \tag{16}$$

A similar calculation for the mean square fluctuation in acceleration $\langle a^2 \rangle = \langle \ddot{x}^2 \rangle$ yields

$$\langle a^2 \rangle = \frac{\hbar \omega_0^3}{2m}. \tag{17}$$

With the above calculations in hand we are now in a position to characterize the ground state of the harmonic oscillator. First, the mean square fluctuation in position given by Eq. (15) matches that obtained in the usual quantum mechanical treatment. Second, the mean square fluctuation in momentum, given by

$$\langle p^2 \rangle = m^2 \langle v^2 \rangle = \frac{m\hbar \omega_0}{2}, \tag{18}$$

also matches that obtained from the standard QM treatment. The harmonic oscillator's ground state energy, kinetic plus potential, is given by

$$\langle E \rangle = \frac{1}{2} m \langle v^2 \rangle + \frac{1}{2} m \omega_0^2 \langle x^2 \rangle = \frac{\hbar \omega_0}{2}, \tag{19}$$

also in agreement with the QM result.

Since the mean position $\langle x \rangle$ and mean momentum $\langle p \rangle$ of the stationary-state oscillator are zero, we also calculate the uncertainty relationship as

$$\Delta x \Delta p = \sqrt{\langle (\Delta x)^2 (\Delta p)^2 \rangle} = \sqrt{\langle (x - \langle x \rangle)^2 \rangle \langle (p - \langle p \rangle)^2 \rangle} = \sqrt{\langle x^2 \rangle \langle p^2 \rangle} = \frac{\hbar}{2}, \tag{20}$$

again in agreement with the known QM result for the harmonic oscillator ground state.

Now, in accordance with the argument being pursued here, we compare the average power being absorbed from the vacuum fluctuation distribution with that radiated due to accelerated motion to determine their relative magnitudes. The power absorbed from the electric field due to the driving force $\mathbf{F} = q\mathbf{E}^{zp}$ is given by

$$\langle P_{abs} \rangle = \langle \mathbf{F} \cdot \mathbf{v} \rangle = \langle q E_x^{zp} \dot{x} \rangle. \tag{21}$$

With $\mathbf{E}^{zp}$ and $\dot{x}$ given by Eqs. (3) and (5), respectively, the calculation carries through as in the derivation of $\langle x^2 \rangle$ above to yield

$$\langle P^{abs} \rangle = \frac{q^2 \hbar \omega_0^3}{12\pi\varepsilon_0 mc^3}. \tag{22}$$

The power radiated due to accelerated motion is given by the standard Larmor expression as [9]

$$\langle P^{rad} \rangle = \frac{q^2 \langle a^2 \rangle}{6\pi\varepsilon_0 c^3} = \frac{q^2 \langle \ddot{x}^2 \rangle}{6\pi\varepsilon_0 c^3}, \tag{23}$$

which, with substitution from Eq. (17) and comparison with Eq. (22), yields

$$\langle P^{rad} \rangle = \langle P^{abs} \rangle. \tag{24}$$

Thus we find that the ground state parameters of the quantum mechanical harmonic oscillator can be accounted for on the basis of interaction between a harmonically-bound point particle and the vacuum electromagnetic zero-point fluctuations. Specifically, the stationary ground state thus established derives from an average balance of power between that absorbed from the vacuum fluctuations and that lost by radiation due to accelerated motion [10]. It can be noted in passing that even in the limit $q \to 0$ (uncharged oscillator) this outcome remains the same as $q^2$ cancels out in the $\langle P^{rad} \rangle = \langle P^{abs} \rangle$ relationship.

## 3 Generalized Approach

Having derived the above relationship between absorbed and radiated powers for the harmonic oscillator's ground state, we now inquire as to whether this balance is specific to the harmonic oscillator by virtue of its simple linear restoring force, or can be extended to more general cases (e.g., nonlinear oscillator, hydrogen atom, particle in a box, etc.).

We begin with the generalization of Eq. (1)

$$m\ddot{\mathbf{r}} = \Gamma m \dddot{\mathbf{r}} + q\left(\mathbf{E}^{zp} + \dot{\mathbf{r}} \times \mathbf{B}^{zp}\right) + \mathbf{F}^{ext}, \tag{25}$$

and we assume $\mathbf{F}^{ext} = -\nabla V$ for a broad class of cases of interest, with $V$ a time-independent confining potential. Multiplication of Eq. (25) by $\dot{\mathbf{r}}$, taking into account mathematical simplifications (e.g., $\dot{\mathbf{r}} \cdot (\dot{\mathbf{r}} \times \mathbf{B}^{zp}) \equiv 0$, $(1/2) d/dt(\dot{\mathbf{r}}^2) = \dot{\mathbf{r}} \cdot \ddot{\mathbf{r}}$, $dV/dt = \partial V/\partial t + \dot{\mathbf{r}} \cdot \nabla V \to \dot{\mathbf{r}} \cdot \nabla V$ for a time-independent potential), followed by collection of terms, leads to

$$\Gamma m \langle \ddot{\mathbf{r}}^2 \rangle + \left\langle \frac{d}{dt}\left( \frac{1}{2} m \dot{\mathbf{r}}^2 + V - \Gamma m \dot{\mathbf{r}} \cdot \ddot{\mathbf{r}} \right) \right\rangle = q \langle \mathbf{E}^{zp} \cdot \dot{\mathbf{r}} \rangle. \qquad (26)$$

For a stationary ground state the second term on the left vanishes and thus the average power radiated due to accelerated motion (Larmor radiation) is balanced by the average power absorbed from the vacuum fluctuations. Substituting the definition of $\Gamma$ from Eq. (8) we obtain

$$\frac{q^2}{6\pi\varepsilon_0 c^3} \langle a^2 \rangle = \langle q \mathbf{E}^{zp} \cdot \mathbf{v} \rangle. \qquad (27)$$

Thus the stationary ground state, although involving accelerated charged-particle motion and hence possessing an associated Larmor radiation loss, is nonetheless observed to be overall radiationless in nature due to the compensatory absorption from the background electromagnetic vacuum zero-point fluctuations. The balance so obtained also accounts for the well-known fact that an oscillator or atom in its ground state does not on net absorb zero-point radiation and therefore remains in its ground state. Finally, we note that this general result is independent of the form of the (time-independent) confining potential $V$ and is thus applicable to a wide range of problems.

**4 Concluding Remarks**

Addressed is the seeming paradox that even though quantum ground states typically involve charged particles in accelerated motions, such states are nonetheless observed to be radiationless in nature. Though this feature is overlooked in everyday application of quantum theory to ground state description, nonetheless this remarkable property is worthy of some discussion and clarification. Such is at hand when one recognizes that ground-state atomic structures are not isolated entities in an empty background, but are perforce immersed in a background of vacuum fluctuations, that of the vacuum electromagnetic zero-point fluctuations being the primary component of interest with regard to the behavior of charged particles. Atoms therefore constitute open systems engaged in dynamic interactions with the surrounding vacuum states. Specifically, the on net radiationless characteristic of the ground state is shown here to derive from particle-vacuum interactions in which a dynamic equilibrium is established between radiation emission due to particle acceleration, and compensatory absorption from the zero-point fluctuations of the vacuum electromagnetic field. Thus, the vacuum field is formally necessary for the stability of atomic structures, and this underlying principle therefore constitutes an important feature of quantum ground states.